\documentclass[review]{elsarticle}

\usepackage{hyperref}
\usepackage{lineno}
\modulolinenumbers[5]

\usepackage{graphicx}
\usepackage{bm}
\hypersetup{colorlinks=false}

\journal{Annals of Physics}

\begin{document}

\begin{frontmatter}

\title{Scalar fields in a five-dimensional Lovelock black hole spacetime}


\author[mymainaddress]{H. S. Vieira\corref{mycorrespondingauthor}}
\cortext[mycorrespondingauthor]{Corresponding author}
\ead{horacio.santana.vieira@hotmail.com}

%\author[mymainaddress]{D. A. T. Vanzella}
%\ead{vanzella@ifsc.usp.br}

%\author[secondaryaddress]{V. B. Bezerra}
%\ead{valdir@fisica.ufpb.br}

\address[mymainaddress]{Instituto de F\'{i}sica de S\~{a}o Carlos, Universidade de S\~{a}o Paulo, Caixa Postal 369, CEP 13560-970, S\~{a}o Carlos, S\~{a}o Paulo, Brazil}
%\address[secondaryaddress]{Departamento de F\'{i}sica, Universidade Federal da Para\'{i}ba, Caixa Postal 5008, CEP 58051-970, Jo\~{a}o Pessoa, PB, Brazil}

\begin{abstract}
We study the interaction between massive scalar fields and the gravitational 
field produced by a higher dimensional Schwarzschild solution in a string cloud model. Exact analytical solutions of both 
angular and radial parts of the Klein--Gordon equation are obtained in terms of 
the three-dimensional spherical harmonics and general Heun functions, 
respectively. From these solutions, we examine the interesting physical phenomena related to the Hawking radiation and resonant frequencies. 
\end{abstract}


%\pacs{02.30.Gp \and 04.20.Jb \and 04.70.-s \and 04.80.Cc \and 47.35.Rs \and 47.90.+a}

\end{frontmatter}

%\linenumbers

%
%%%%%%%%%%%%%%%%%%%%%%%%%%%%%%%%%%%%%%%%%%%%%%%%%%%%%%%%%%%%%%%%%%%%%%%%%%%%%%%%%%%%%%%%%%%%%% Introduction
%
\section{Introduction}

The theory that combines quantum mechanics and general relativity is still
incomplete. Thus, many investigations have been done in order to obtain some answers
that can be used to construct the expected quantum gravity
theory~\cite{PhysRevD.18.1747,PhysRevD.70.044025,JPhysAMathTheor.45.244004,PhysRevD.92.121501(R),EurPhysJC.77.52}.
Among these lines of research, we are specially interested in the one that
investigates the interaction between quantum fields and black hole spacetimes. In
particular, when scalar fields are considered, a lot of results can be found in the
literature, as for example, on some as aspects of the quasinormal modes and
quantization~\cite{GenRelativGravit.50.125}, the scattering
waves~\cite{PhysLettB.786.300}, the tunneling of scalar
particles~\cite{EurPhysJC.79.692}, the black hole hair from scalar dark
matter~\cite{JCAP.06.038}, and about the growth of massive scalar hair around a
Schwarzschild black hole~\cite{PhysRevD.100.063014}.

Among the various physical phenomena due to the presence of a black hole in a
spacetime region, we are particularly interested in investigating the ones related to
the black hole radiation and the quasispectrum of scalar particles. Black holes may
emit a thermal spectrum related to the quantum mechanical effects on their exterior
event horizons~\cite{Nature.248.30,CommunMathPhys.43.199}. On the other hand, scalar
fields may present an oscillating energy spectrum (quasispectrum) after the
interaction with a black
hole~\cite{PhysRevD.81.104042,PhysRevD.90.124071,AdvHighEnergyPhys.2015.739153,PhysRevD.94.084040,ChinPhysC.43.035102,IntJModPhysD.28.1950151}.

In this work we obtain the exact solutions of the Klein--Gordon equation and then investigate the Hawking radiation and the resonant frequencies of massive scalar fields in the five-dimensional Schwarzschild black hole in a string cloud model, which is a Lovelock class of black hole.

In the early 1970s, motivated by the fact that the general relativity cannot be
quantized, it was proposed some modified theories of gravity, for example, the
$f(R)$ gravity~\cite{MonNotRoyAstronSoc.150.1} and the Lovelock
theory~\cite{JMathPhys.12.498}, among others. In this scenario, we will deal with a
black hole spacetime solution of the higher curvature gravity theory, namely, the
so-called Lovelock gravity, which is free from ghosts and hence it is a natural
extension to Einstein gravity. The Lovelock theory of gravity is known as the most
general mathematically consistent metric theory leading to second order equations of
motion in arbitrary number of spacetime dimensions~\cite{JCAP.05.050}.

On the other hand, in the case of a five-dimensional spacetime, Herscovich
\textit{et al.}~\cite{PhysLettB.689.192} obtained a black hole solution in the
Einstein--Gauss--Bonnet theory for the string cloud model and explored the
thermodynamical global and local stability of the system with vanishing or
non-vanishing cosmological constant.

In the present paper, the radial solution of the Klein--Gordon equation will be given
in terms of the general Heun functions. These special functions of mathematical
physics have gained increasingly more importance due to their large number of
applications in different areas of natural science, from biology to
physics~\cite{AdvHighEnergyPhys.2018.8621573}. Here, we will impose some boundary
conditions to these functions in order to study the both Hawking radiation and
resonant frequencies.

This paper is organized as follows. In Section 2, we present the Lovelock solution for a five-dimensional spherical symmetric black hole spacetime. In Section 3, we solve the Klein--Gordon equation in the background under consideration. In Section 4, we examine the Hawking radiation. Section 5, we discuss the resonant frequencies of massive scalar particles. Finally, in Section 6, we present our conclusions.
%
%%%%%%%%%%%%%%%%%%%%%%%%%%%%%%%%%%%%%%%%%%%%%%%%%%%%%%%%%%%%%%%%%%%%%%%%%%%%%%%%%%%%%%%%%%%%%% Lovelock black hole solution
%
\section{Lovelock black hole solution}\label{LBH}
When the Einstein's general relativity is generalized in higher dimensions by keeping
almost its characteristics, we obtain the so-called Lovelock gravity, which gives
second order field equations in arbitrary dimensions. The Lovelock action appears as
the low energy limit of a heterotic superstring theory~\cite{PhysRevD.89.084027}. In
this approach, the higher dimensional Schwarzschild solution in a string cloud model
is given by
\begin{equation}
ds^{2}=g_{\sigma\tau}dx^{\sigma}dx^{\tau}=-f(r)\ dt^{2}+\frac{1}{f(r)}\ dr^{2}+r^{2}\ d\Omega^{2}_{D-2},
\label{eq:metrica_Lovelock}
\end{equation}
with
\begin{equation}
f(r)=1-\frac{2M}{(D-3)r^{D-3}}-\frac{2a}{(D-2)r^{D-4}},
\label{eq:f_Lovelock}
\end{equation}
%\xpar\cvskip[-11pt] 
\begin{equation}
d\Omega^{2}_{D-2}=\tilde{\gamma}_{ij}dx^{i}dx^{j},
\label{eq:angulo_solido_Lovelock}
\end{equation}
where $D$ is the spacetime dimension, $M$ represents a spherical mass
centered at the origin of the system of coordinates, and $a$ is a constant
related to the energy momentum tensor of a string cloud. The line element $d\Omega^{2}_{D-2}$
is related to a $(D-2)$-dimensional hypersurface with constant curvature
$k=-1,0$ or $+1$, which corresponds to the hyperbolic, flat or spherical
spaces, respectively. It is worth calling attention to the fact that for $D=4$
the Letelier black hole spacetime~\cite{PhysRevD.20.1294} is recovered, with an event
horizon of radius $r_{s}=2M/(1-a)$. Note that we adopted the natural units where
$G \equiv c \equiv \hbar \equiv 1$.

In this work, we will focus on a five-dimensional spherically symmetric spacetime, so that
\begin{equation}
d\Omega^{2}_{3}=d\phi^{2}+\sin^{2}\phi(d\theta^{2}+\sin^{2}\theta\ d\varphi^{2}).
\label{eq:angulo_solido_Lovelock_D5}
\end{equation}
This is the line element of a three-sphere, where $\phi$ and $\theta$ run over the range 0 to $\pi$, and $\varphi$ runs over 0 to $2\pi$. Thus, the metric of a Lovelock black hole spacetime in five-dimensions can be written as
\begin{equation}
ds^{2}=-f(r)\ dt^{2}+\frac{1}{f(r)}\ dr^{2}+r^{2}[d\phi^{2}+\sin^{2}\phi(d\theta^{2}+\sin^{2}\theta\ d\varphi^{2})],
\label{eq:metrica_Lovelock_D5}
\end{equation}
with
\begin{equation}
f(r)=1-\frac{M}{r^2}-\frac{2a}{3r}.
\label{eq:f_Lovelock_D5}
\end{equation}
The background under consideration has the apparent horizon ($\mbox{\tiny{AH}}$), which is the outermost marginally trapped surface for the outgoing photons, given by the zeros of $f(r)=0$, that is,
\begin{equation}
f(r)=0=(r-r_{+})(r-r_{-}).
\label{eq:f_surface_Lovelock_D5}
\end{equation}
The solutions of this equation are given by
\begin{equation}
r_{\mbox{\tiny{AH}}} = \frac{1}{3}(a \pm \sqrt{a^{2}+9M}).
\label{eq:horizons_Lovelock_D5}
\end{equation}
For simplicity and convenience, we will refer to these solutions as the exterior
($r_{+}$) and interior ($r_{-}$) apparent horizons. In fact, the photon can
escape from the apparent horizon and reach an arbitrary large distance, which will
confirm that the surface located at $r=r_{\mbox{\tiny{AH}}}$ is an apparent horizon not an event
horizon. Furthermore, it is worth calling attention to the fact that the cloud of
strings alone can have an apparent horizon located at $r_{\mbox{\tiny{AH}}}=2a/3$ (for a detailed
review about the radiating black hole horizons, see~\cite{PhysRevD.89.084027} and
references therein).

In what follows we will consider massive scalar fields propagating in this back\-ground.
%
%%%%%%%%%%%%%%%%%%%%%%%%%%%%%%%%%%%%%%%%%%%%%%%%%%%%%%%%%%%%%%%%%%%%%%%%%%%%%%%%%%%%%%%%%%%%%% Klein-Gordon equation
%
\section{Klein-Gordon equation}\label{KG}
We want to investigate the behavior of massive scalar fields interacting with the five-dimensional Schwarzschild black hole in a string cloud model given by Eq.~\ref{eq:metrica_Lovelock_D5}. To do this, we need to solve the Klein--Gordon equation, which is given by
\begin{equation}
\biggl[\frac{1}{\sqrt{-g}}\partial_{\sigma}(g^{\sigma\tau}\sqrt{-g}\partial_{\tau})-\mu^{2}\biggr]\Psi=0,
\label{eq:Klein-Gordon_gauge_Lovelock_D5}
\end{equation}
where $\mu$ is the mass of the scalar particle.

Thus, substituting Eq.~\ref{eq:metrica_Lovelock_D5} into Eq.~\ref{eq:Klein-Gordon_gauge_Lovelock_D5}, we obtain
\begin{equation}
\biggl\{-\frac{r^{2}}{f(r)}\frac{\partial^{2}}{\partial t^{2}}+\frac{1}{r}\frac{\partial}{\partial r}\biggl[r^{3}f(r)\frac{\partial}{\partial r}\biggr]-\mathbf{L}^{2}_{\phi\theta\varphi}-r^{2}\mu^{2}\biggr\}\Psi=0,
\label{eq:mov_1_Lovelock_D5}
\end{equation}
where $\mathbf{L}^{2}_{\phi\theta\varphi}$ is the three-dimensional angular momentum operator, which is given by
\begin{equation}
\mathbf{L}^{2}_{\phi\theta\varphi}=-\frac{1}{\sin^{2}\phi}\frac{\partial}{\partial \phi}\biggl(\sin^{2}\phi \frac{\partial}{\partial \phi}\biggr)+\frac{1}{\sin^{2}\phi}\mathbf{L}^{2},
\label{eq:angular_operator_Lovelock_D5}
\end{equation}
with
\begin{equation}
\mathbf{L}^{2}=-\frac{1}{\sin\theta}\frac{\partial}{\partial \theta}\biggl(\sin\theta\frac{\partial}{\partial \theta}\biggr)-\frac{1}{\sin^{2}\theta}\frac{\partial^{2}}{\partial\varphi^{2}}
\label{eq:angular_operator}
\end{equation}
being the well known angular momentum operator.

Now, we need to establish the form of the scalar wave function $\Psi$. Since that the spacetime under consideration is static and spherically symmetric, the scalar wave function can be written as%\xlooseness[-1]
\begin{equation}
\Psi(\mathbf{r},t)=R(r)Y_{slm}(\phi,\theta,\varphi)\mbox{e}^{-i \omega t},
\label{eq:separacao_variaveis_Lovelock_D5}
\end{equation}
where $\omega$ is the frequency (energy, because we are considering $\hbar=1$).
Note that the general angular solution is given in terms of the three-dimensional
normalized spherical harmonic function $Y_{slm}(\phi,\theta,\varphi)=P_{s,4}^{l}(\cos\phi)Y_{lm}(\theta,\phi)$, where $P_{s,4}^{l}(\cos\phi)$ is the
associated Legendre function in
four-dimensions~\cite{RevMexFis.59.248,Spherical:2014}. Therefore, substituting
Eq.~\ref{eq:separacao_variaveis_Lovelock_D5} into
Eq.~\ref{eq:mov_1_Lovelock_D5}, we obtain the following radial equation
\begin{equation}
\frac{1}{r}\frac{d}{dr}\biggl[r^{3}f(r)\frac{dR}{dr}\biggr]+\biggl[\frac{r^{2}\omega^{2}}{f(r)}-(\lambda_{slm}+r^{2}\mu^{2})\biggr]R=0,
\label{eq:mov_radial_1_Lovelock_D5}
\end{equation}
where we have used the fact that $\mathbf{L}^{2}_{\phi\theta\varphi}Y_{slm}(\phi,\theta,\varphi)=\lambda_{slm}Y_{slm}(\phi,\theta,\varphi)$, with $\lambda_{slm}=s(s+2)$.
%
%%%%%%%%%%%%%%%%%%%%%%%%%%%%%%%%%%%%%%%%%%%%%%%%%%%%%%%%%%%%%%%%%%%%%%%%%%%%%%%%%%%%%%%%%%%%%% Radial solution
%
\subsection{Radial solution}
Let us solve the radial part of the Klein--Gordon equation in the five-di\-men\-sion\-al Schwarzschild black hole in a string cloud model. First, we use Eq.~\ref{eq:f_Lovelock_D5} and write down Eq.~\ref{eq:mov_radial_1_Lovelock_D5} as
\begin{eqnarray}
&& \frac{d^{2}R}{dr^{2}}+\biggl(\frac{3}{r}+\frac{1}{r-r_{+}}+\frac{1}{r-r_{-}}\biggr)\frac{dR}{dr}\nonumber\\
&&+  \frac{1}{(r-r_{+})(r-r_{-})}\biggl[\frac{\omega^{2}}{(r-r_{+})(r-r_{-})}-(\lambda_{slm}+r^{2}\mu^{2})\biggr]R=0.
\label{eq:mov_radial_2_Lovelock_D5}
\end{eqnarray}
Note that this equation has singularities at $r=(0,r_{+},r_{-},\infty)$. Then, the transformation of Eq.~\ref{eq:mov_radial_2_Lovelock_D5} to a Heun-type equation is achieved by setting a new radial coordinate $x$ as
\begin{equation}
x=\frac{r-r_{+}}{r_{-}-r_{+}},
\label{eq:homog_subs_radial_2_Lovelock_D5_x}
\end{equation}
which transforms $(r_{+},r_{-}) \mapsto (0,1)$. In addition, the remaining singularity is transformed to $x=b$, where
\begin{equation}
b=\frac{-r_{+}}{r_{-}-r_{+}}.
\label{eq:singularity_b_2_Lovelock_D5}
\end{equation}
It is worth calling attention to the fact that this is an homographic substitution of the independent variable, which has the following asymptotic regimes: $x \rightarrow 0 \Rightarrow r \rightarrow r_{+}$ and $x \rightarrow \infty \Rightarrow r \rightarrow \infty$. Therefore, it is easy to see that this transformation covers the entire spacetime region of the five-dimensional Schwarzschild black hole in a string cloud model. Thus, substituting Eq.~\ref{eq:homog_subs_radial_2_Lovelock_D5_x} into Eq.~\ref{eq:mov_radial_2_Lovelock_D5}, we obtain
\begin{eqnarray}
&& \frac{d^{2}R}{dx^{2}}+\biggl(\frac{1}{x}+\frac{1}{x-1}+\frac{3}{x-b}\biggr)\frac{dR}{dx}\nonumber\\
&&+  \biggl\{\frac{b^2 \omega ^2}{r_{+}^2 }\frac{1}{x^2}+\frac{b^2 \omega ^2}{r_{+}^2 }\frac{1}{(x-1)^2}-\frac{b \lambda_{slm} }{(b-1) r_{+}^2 }\frac{1}{(x-b)^2}\nonumber\\
&&+  \frac{2 b^2 \omega ^2+\lambda_{slm} +\mu ^2 r_{+}^2}{r_{+}^2 }\frac{1}{x}\nonumber\\
&&+  \frac{-2 b^4 \omega ^2+4 b^3 \omega ^2-b^2 \lambda_{slm} -b^2 \mu ^2 r_{+}^2-2 b^2 \omega ^2+2 b \mu ^2 r_{+}^2-\mu ^2 r_{+}^2}{(b-1)^2 r_{+}^2 }\frac{1}{x-1}\nonumber\\
&&+  \frac{(2 b-1) \lambda_{slm} }{(b-1)^2 r_{+}^2 }\frac{1}{x-b}\biggr\}R=0.
\label{eq:mov_radial_2_Lovelock_D5_x}
\end{eqnarray}

Now, let us perform a transformation in order to reduce the powers of the terms proportional to $1/x^{2}$, $1/(x-1)^{2}$, and $1/(x-b)^{2}$. This transformation is a F-homotopic transformation of the dependent variable, $R(x) \mapsto U(x)$, such that
\begin{equation}
R(x)=x^{A_{1}}(x-1)^{A_{2}}(x-a)^{A_{3}}U(x),
\label{eq:F-homotopic_mov_radial_2_Lovelock_D5_x}
\end{equation}
where the coefficients $A_{1}$, $A_{2}$ and $A_{3}$ are given by
\begin{equation}
A_{1}=\frac{i b \omega }{r_{+}},
\label{eq:A1_radial_2_Lovelock_D5_x}
\end{equation}
%\xpar\cvskip[-8pt] 
\begin{equation}
A_{2}=\frac{i b \omega }{r_{+}},
\label{eq:A2_radial_2_Lovelock_D5_x}
\end{equation}
%\xpar\cvskip[-8pt] 
\begin{equation}
A_{3}=\frac{\sqrt{r_{+}^2 (b-1)^2 +b \lambda_{slm}(b-1)}- r_{+}(b-1)}{r_{+}(b-1)}.
\label{eq:A3_radial_2_Lovelock_D5_x}
\end{equation}
In this case, the new radial function $U(x)$ satisfies the following equation
\begin{equation}
\frac{d^{2}U}{dx^{2}}+\biggl(\frac{1+2A_{1}}{x}+\frac{1+2A_{2}}{x-1}+\frac{3+2A_{3}}{x-b}\biggr)\frac{dU}{dx}+\frac{A_{4}x-A_{5}}{x(x-1)(x-b)}U=0,
\label{eq:mov_radial_2_Lovelock_D5_x_U}
\end{equation}
where the coefficients $A_{4}$ and $A_{5}$ are given by
\begin{eqnarray}
A_{4} & = & \frac{1}{(b-1) r_{+}^2} [2 A_{1} (b-1) r_{+}^2 (A_{2}+A_{3}+2)+2 A_{2} (A_{3}+2) (b-1) r_{+}^2\nonumber\\
&&+  2 A_{3} b r_{+}^2-2 A_{3} r_{+}^2-2 b^3 \omega ^2+2 b^2 \omega ^2+b \lambda_{slm} -b \mu ^2 r_{+}^2+\mu ^2 r_{+}^2],
\label{eq:A4_radial_2_Lovelock_D5_x}
\end{eqnarray}
%\xpar\cvskip[-11pt]
\begin{equation}
A_{5}=A_{1} (2 A_{2} b+2 A_{3}+b+3)+A_{3}-\frac{b \left(-A_{2} r_{+}^2+2 b^2 \omega ^2+\lambda_{slm} +\mu ^2 r_{+}^2\right)}{r_{+}^2}.
\label{eq:A5_radial_2_Lovelock_D5_x}
\end{equation}
Thus, Eq.~\ref{eq:mov_radial_2_Lovelock_D5_x_U} is similar to the general Heun
equation~\cite{Ronveaux:1995}, which is a Fuchsian type equation with four regular
singularities located at $x=(0,1,b,\infty)$. The canonical form of the general Heun equation
is written as
\begin{equation}
\frac{d^{2}U}{dx^{2}}+\biggl(\frac{\gamma}{x}+\frac{\delta}{x-1}+\frac{\epsilon}{x-b}\biggr)\frac{dU}{dx}+\frac{\alpha\beta x-q}{x(x-1)(x-b)}U=0,
\label{eq:canonical_form_general_Heun}
\end{equation}
where $U(x)=\mbox{HeunG}(b,q;\alpha,\beta,\gamma,\delta;x)$ is the general Heun function, which is simul\-tane\-ous\-ly a local Frobenius solution around a singularity $x=b_{i}$ and a local Frobenius solution around $x=b_{j}$, so that it is analytic in some domain including both these singularities. The singularity parameter $b$ is such that $b \neq 0,1$. The parameters $\alpha$, $\beta$, $\gamma$, $\delta$, $\epsilon$, $q$ and $b$ are generally complex, arbitrary, and related by $\gamma+\delta+\epsilon=\alpha+\beta+1$.

Therefore, the exact analytical solution of the radial part of the Klein--Gordon equation, for a massive scalar particle propagating in the five-di\-men\-sion\-al Schwarzschild black hole spacetime in a string cloud model, is given by
\begin{eqnarray}
R(x) & = & x^{\frac{1}{2}(\gamma-1)}(x-1)^{\frac{1}{2}(\delta-1)}(x-a)^{\frac{1}{2}(\epsilon-3)}\nonumber\\
&&\times  \{C_{1}\ \mbox{HeunG}(b,q;\alpha,\beta,\gamma,\delta;x)\nonumber\\
&&+  C_{2}\ x^{1-\gamma}\ \mbox{HeunG}(b,q_{1};\alpha_{1},\beta_{1},\gamma_{1},\delta;x)\},
\label{eq:general_solution_radial_2_Lovelock_D5_x}
\end{eqnarray}
where $C_{1}$ and $C_{2}$ are constants to be determined. The parameters $\alpha$, $\beta$, $\gamma$, $\delta$, $\epsilon$ and $q$ are given by
\begin{equation}
\alpha=1+\frac{2 i b \omega}{r_{+}}+(4+\mu^2)^{\frac{1}{2}}+\biggl[1+\frac{b \lambda_{slm}}{(b-1) r_{+}^2}\biggr]^{\frac{1}{2}},
\label{eq:alpha_radial_2_Lovelock_D5_x}
\end{equation}
%\xpar\cvskip[-8pt]
\begin{equation}
\beta=\alpha-2(4+\mu^2)^{\frac{1}{2}},
\label{eq:beta_radial_2_Lovelock_D5_x}
\end{equation}
%\xpar\cvskip[-8pt]
\begin{equation}
\gamma=1+\frac{2 i \omega }{r_{+}-r_{-}},
\label{eq:gamma_radial_2_Lovelock_D5_x}
\end{equation}
%\xpar\cvskip[-8pt]
\begin{equation}
\delta=1+\frac{2 i \omega }{r_{+}-r_{-}},
\label{eq:delta_radial_2_Lovelock_D5_x}
\end{equation}
%\xpar\cvskip[-8pt]
\begin{equation}
\epsilon=\frac{2 \sqrt{(b-1) r_{+}^2 [b \lambda_{slm} +(b-1) r_{+}^2]}+(b-1) r_{+}^2}{(b-1) r_{+}^2},
\label{eq:eta_radial_2_Lovelock_D5_x}
\end{equation}
%\xpar\cvskip[-8pt]
\begin{eqnarray}
q & = & \frac{i b (2 b^2-b-1) r_{+}^2 \omega-r_{+}^3 [(b-1)(b \mu ^2+1)]}{(b-1) r_{+}^3}\nonumber\\
%\QUERY[4]
&&+  \frac{\sqrt{(b-1) r_{+}^2 [b \lambda_{slm} +(b-1) r_{+}^2]}-(b-1)(4 b^3 \omega ^2+b \lambda_{slm})}{(b-1) r_{+}^2}\nonumber\\
&&+  \frac{2 i b \omega  \sqrt{(b-1) r_{+}^2 [b \lambda_{slm} +(b-1) r_{+}^2]}}{(b-1) r_{+}^3}.
\label{eq:q_radial_2_Lovelock_D5_x}
\end{eqnarray}
Furthermore, the parameters $\alpha_{1}$, $\beta_{1}$, $\gamma_{1}$ and $q_{1}$ are given by
\begin{equation}
\alpha_{1}=\alpha+1-\gamma,
\label{eq:alpha_1_general_Heun}
\end{equation}
%\xpar\cvskip[-11pt]
\begin{equation}
\beta_{1}=\beta+1-\gamma,
\label{eq:beta_1_general_Heun}
\end{equation}
%\xpar\cvskip[-11pt]
\begin{equation}
\gamma_{1}=2-\gamma.
\label{eq:gamma_1_general_Heun}
\end{equation}
%\xpar\cvskip[-11pt]
\begin{equation}
q_{1}=q+(\alpha\delta+\epsilon)(1-\gamma).
\label{eq:q_1_general_Heun}
\end{equation}

Next, we will use this radial solution to discuss some interesting physical phe\-nom\-e\-na, namely, the Hawking radiation and the resonant frequencies.
%
%%%%%%%%%%%%%%%%%%%%%%%%%%%%%%%%%%%%%%%%%%%%%%%%%%%%%%%%%%%%%%%%%%%%%%%%%%%%%%%%%%%%%%%%%%%%%% Hawking radiation
%
\section{Hawking radiation}\label{Hawking}
In this section we will examine the Hawking radiation of massive scalar particles near to the exterior apparent horizon of a five-dimensional Schwarzschild black hole spacetime in a string cloud model. Mathematically, this means that we need to impose the limit $r \rightarrow r_{+}$ to the radial solution.

In this limit, from Eq.~\ref{eq:homog_subs_radial_2_Lovelock_D5_x}, we have that
$x \rightarrow 0$. Thus, we must first know how the general Heun functions behave near the
point $x=0$. If $\gamma \neq 0,-1,-2,\ldots$, the general Heun function is analytic in the disk
$|x|  < 1$, and has the following Maclaurin expansion~\cite{MathAnn.33.161}
\begin{equation}
\mbox{HeunG}(b,q;\alpha,\beta,\gamma,\delta;x)=\sum_{j=0}^{\infty}c_{j}x^{j},
\label{eq:serie_HeunG_todo_x}
\end{equation}
where
\begin{eqnarray}
 b\gamma c_{1}-qc_{0}=0,&&\nonumber\\
 X_{j}c_{j+1}-(Q_{j}+q)c_{j}+P_{j}c_{j-1}=0, \quad j \geq 1,&&
\label{eq:recursion_General_Heun}
\end{eqnarray}
with $c_{0}=1$ and
\begin{eqnarray}
 P_{j}=(j-1+\alpha)(j-1+\beta),&&\nonumber\\
 Q_{j}=j[(j-1+\gamma)(1+b)+b\delta+\epsilon],&&\nonumber\\
 X_{j}=b(j+1)(j+\gamma).&&
\label{eq:P_Q_X_recursion_General_Heun}
\end{eqnarray}
Then, we can conclude that $\mbox{HeunG}(b,q;\alpha,\beta,\gamma,\delta;x \rightarrow 0) \sim 1$.

Now, by imposing the limit $r \rightarrow r_{+}$ to the radial solution given by Eq.~\ref{eq:general_solution_radial_2_Lovelock_D5_x}, we obtain the following asymptotic behavior
\begin{equation}
R(r) \sim C_{3}\ (r-r_{+})^{\frac{1}{2}(\gamma-1)}+C_{4}\ (r-r_{+})^{-\frac{1}{2}(\gamma-1)},
\label{eq:exp_0_solucao_general_solution_radial_2_Lovelock_D5_x}
\end{equation}
where all constants are included in $C_{3}$ and $C_{4}$. Then, taking into account the solution of the time dependence, we can write the wave solution as
\begin{equation}
\Psi=\mbox{e}^{-i \omega t}(r-r_{+})^{\pm\frac{1}{2}(\gamma-1)}.
\label{eq:sol_onda_radial_2_Lovelock_D5_x}
\end{equation}
From Eq.~\ref{eq:gamma_radial_2_Lovelock_D5_x}, for the parameter $\gamma$, we obtain
\begin{equation}
\frac{1}{2}(\gamma-1)=\frac{i}{2\kappa_{+}}\omega,
\label{eq:beta/2_solucao_geral_radial_2_Lovelock_D5_x}
\end{equation}
where $\kappa_{+}$ is the gravitational acceleration on the background horizon surface $r_{+}$ given by
\begin{equation}
\kappa_{+} \equiv \frac{1}{2}\left.\frac{df(r)}{dr}\right|_{r=r_{+}}=\frac{r_{+}-r_{-}}{2}.
\label{eq:acel_grav_ext_Lovelock_D5}
\end{equation}
Thus, on the five-dimensional Schwarzschild black hole exterior horizon surface, the ingoing and outgoing wave solutions are given by
\begin{equation}
\Psi_{in}=\mbox{e}^{-i \omega t}(r-r_{+})^{-\frac{i}{2\kappa_{+}}\omega},
\label{eq:sol_in_1_Lovelock_D5}
\end{equation}
%\xpar\cvskip[-11pt]
\begin{equation}
\Psi_{out}(r>r_{+})=\mbox{e}^{-i \omega t}(r-r_{+})^{\frac{i}{2\kappa_{+}}\omega}.
\label{eq:sol_out_2_Lovelock_D5}
\end{equation}

Now, we follow the approach developed by Vieira et \textit{al.}~\cite{AnnPhys.350.14}
in order to obtain a real damped part of the outgoing wave solution of the scalar
field, which can be used to construct an explicit expression for the decay rate
$\Gamma$. Thus, the relative scattering probability of the scalar wave at the
exterior apparent horizon surface $r=r_{+}$ is given by
\begin{equation}
\Gamma_{+}=\left|\frac{\Psi_{out}(r>r_{+})}{\Psi_{out}(r<r_{+})}\right|^{2}=\mbox{e}^{-\frac{2\pi}{\kappa_{+}}\omega}.
\label{eq:taxa_refl_Lovelock_D5}
\end{equation}

Finally, by using the Damour--Ruffini--Sannan
method~\cite{PhysRevD.14.332,GenRelativGravit.20.239}, we get the resulting Hawking
radiation spectrum of massive scalar particles, which is given by
\begin{equation}
\bar{N}_{\omega}=\frac{\Gamma_{+}}{1-\Gamma_{+}}=\frac{1}{\mbox{e}^{\frac{2\pi}{\kappa_{+}}\omega}-1}.
\label{eq:espectro_rad_Lovelock_D5_2}
\end{equation}
Therefore, we can see that the resulting Hawking radiation spectrum of massive scalar particles in the five-dimensional Schwarzschild black hole spacetime in a string cloud model has a thermal character, analogous to the black body spectrum, where $k_{B}T_{+}=\hbar\kappa_{+}/2\pi$. This result shows that the Hawking radiation occurs even in high dimensions, that is, it is really a phenomenon due to the effective geometry.
%
%%%%%%%%%%%%%%%%%%%%%%%%%%%%%%%%%%%%%%%%%%%%%%%%%%%%%%%%%%%%%%%%%%%%%%%%%%%%%%%%%%%%%%%%%%%%%% Resonant frequencies
%
\section{Resonant frequencies}\label{RFs}
In this section we will investigate a kind of quasispectrum, the so-called resonant
frequencies, which are related to the decay of the perturbation field, that is, they
correspond to damped oscillations. Mathematically, this means that we need to impose
a boundary condition to the radial solution, namely, it should be well behaved at
asymptotic infinity~\cite{AnnPhys.373.28}.

In this limit, we have that $R(x)$ should be a polynomial. Thus, we must first
know how the general Heun functions becomes a polynomial. Indeed, the function
$\mbox{HeunG}(a,q;\alpha,\beta,\gamma,\delta;x)$ turns to be a polynomial of degree $n$ if it satisfies the
$\alpha$-condition, which is given by~\cite{Ronveaux:1995}
\begin{equation}
\alpha=-n,
\label{eq:condiction_poly_General_Heun}
\end{equation}
where $n=0,1,2,\ldots$ is a quantum number.

Now, by imposing the $\alpha$-condition to the radial solution given by Eq.~\ref{eq:general_solution_radial_2_Lovelock_D5_x}, we obtain the following expression for the resonant frequencies
\begin{equation}
\omega_{n}=i\frac{\sqrt{a^{2}+9M}}{3}\biggl(n+1+\sqrt{4+\mu^{2}}+\sqrt{\frac{M-\lambda_{slm}}{M}}\biggr).
\label{eq:RFs_Lovelock_D5}
\end{equation}
Note that this quasispectrum is a complex number with just the imaginary part and hence it corresponds to the decay rate of the oscillation. In addition, these eigenvalues are degenerate, because there is a dependence on the separation constant $\lambda_{slm}=s(s+2)$.

It is worth to calling attention to the fact that we have obtained the massive scalar resonant frequencies in the five-dimensional Schwarzschild black hole spacetime in a string cloud model directly from the general Heun function by using the condition which should be
imposed in such a way that this function reduces to a polynomial, and that there is no similar result in the literature for this case.

\begin{figure}
\caption{The scalar resonant frequencies of a five-dimensional Schwarzschild black hole as a function of $a$ for $M=10$, $\mu=1$ and $n=0$.\label{fig:Fig1_Lovelock_D5}}
\includegraphics{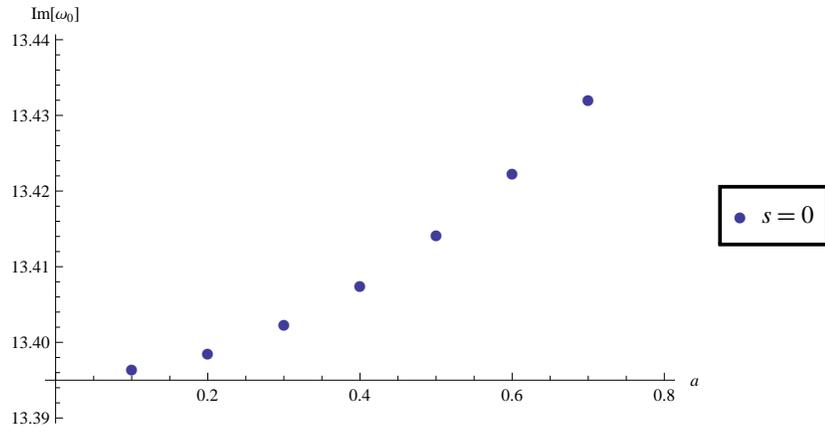}
\end{figure}

\begin{figure}
\caption{The scalar resonant frequencies of a five-dimensional Schwarzschild black hole as a function of $a$ for $M=10$, $\mu=1$ and $n=2$.\label{fig:Fig2_Lovelock_D5}}
\includegraphics{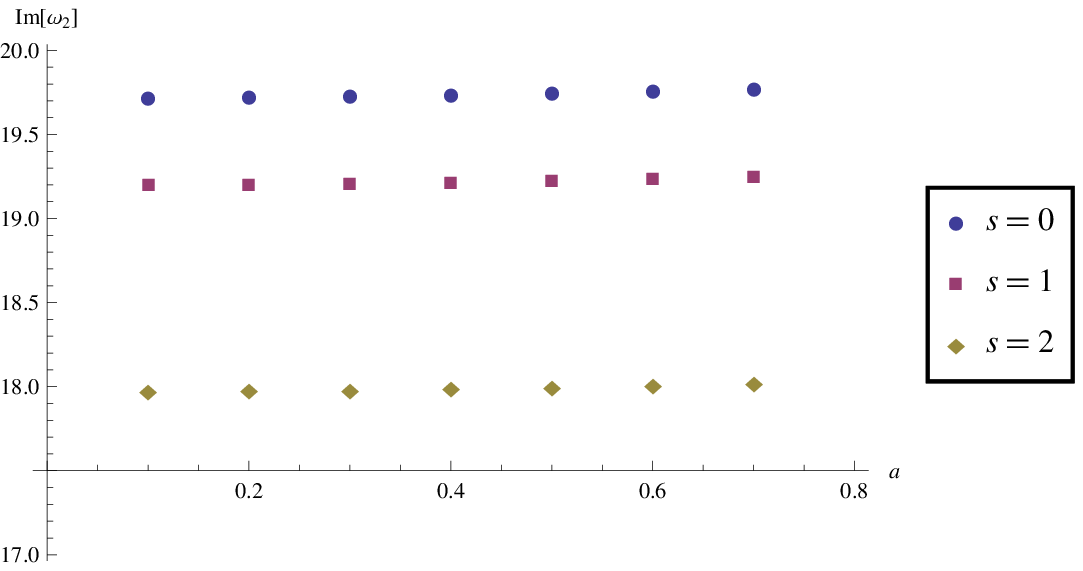}
\end{figure}

\begin{figure}
\caption{The scalar resonant frequencies of a five-dimensional Schwarzschild black hole as a function of $a$ for $M=10$, $\mu=1$ and $s=0$.\label{fig:Fig3_Lovelock_D5}}
\includegraphics{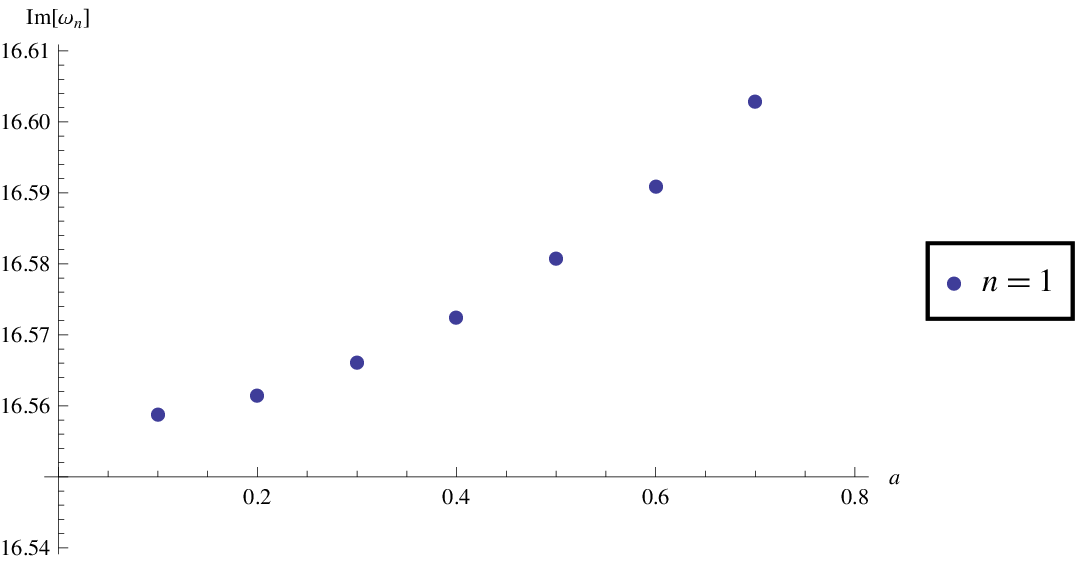}
\end{figure}

\begin{figure}
\caption{The scalar resonant frequencies of a five-dimensional Schwarzschild black hole as a function of $a$ for $M=10$, $\mu=1$ and $s=0$.\label{fig:Fig4_Lovelock_D5}}
\includegraphics{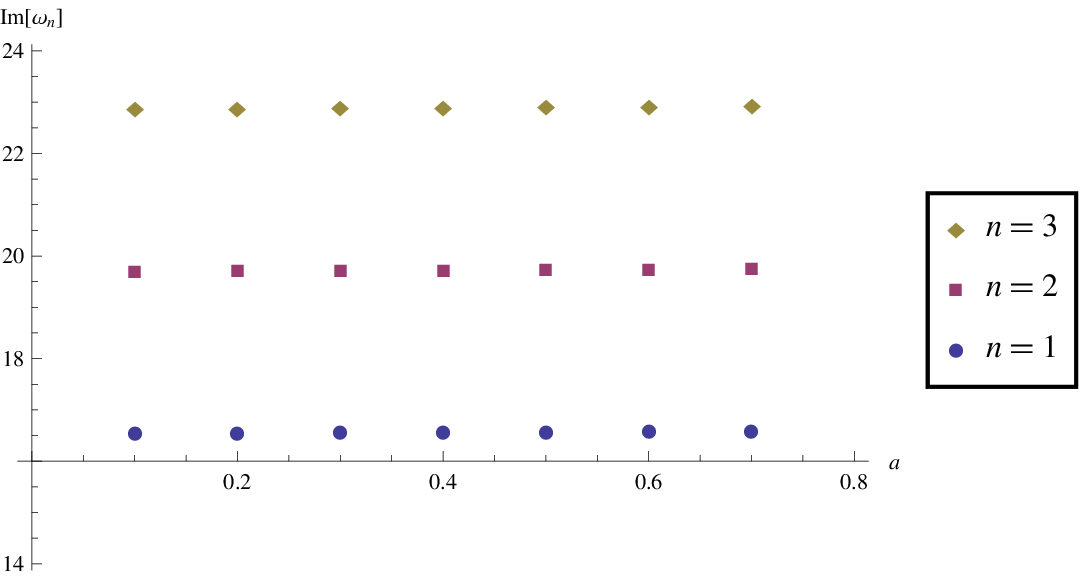}
\end{figure}

The resonant frequencies given by Eq.~\ref{eq:RFs_Lovelock_D5} are shown in Figs.~\ref{fig:Fig1_Lovelock_D5}-\ref{fig:Fig4_Lovelock_D5} as functions of the parameter $a$.

In Fig.~\ref{fig:Fig1_Lovelock_D5} we see that for a fixed value of the eigenvalue $s$, the (imaginary part of) resonant frequencies increase with $a$. The same can be concluded from Fig.~\ref{fig:Fig3_Lovelock_D5}, with a fixed value of the quantum number $n$.

From Fig.~\ref{fig:Fig2_Lovelock_D5} we conclude that the (imaginary part of) resonant frequencies decrease with $s$ for fixed values of the parameter $a$. The opposite occurs in Fig.~\ref{fig:Fig4_Lovelock_D5}, with fixed values of the quantum number $n$.

The resonant frequencies given by Eq.~\ref{eq:RFs_Lovelock_D5} are also shown in Tables \ref{tab:Tab1_Lovelock_D5}-\ref{tab:Tab4_Lovelock_D5} as functions of $a$, $\mu$, $s$, and $n$, respectively.

\begin{table}[!ht]
\caption{The scalar resonant frequencies five-dimensional Schwarzschild black hole for $M=10$, $\mu=1$ and $s=0$.}
\label{tab:Tab1_Lovelock_D5}
\begin{tabular}{ccc}
\hline\noalign{\smallskip}
			$a$  & $\mbox{Re}[\omega_{0}]$ & $\mbox{Im}[\omega_{0}]$ \\
\noalign{\smallskip}\hline\noalign{\smallskip}
			0.0 & 0.00000 & 13.39562 \\
			0.1 & 0.00000 & 13.39637 \\
			0.2 & 0.00000 & 13.39860 \\
			0.3 & 0.00000 & 13.40232 \\
			0.4 & 0.00000 & 13.40753 \\
			0.5 & 0.00000 & 13.41422 \\
			0.6 & 0.00000 & 13.42239 \\
			0.7 & 0.00000 & 13.43204 \\
\noalign{\smallskip}\hline
\end{tabular}
\end{table}

\begin{table}[!ht]
\caption{The scalar resonant frequencies five-dimensional Schwarzschild black hole for $M=10$, $a=0.1$ and $s=1$.}
\label{tab:Tab2_Lovelock_D5}
\begin{tabular}{ccc}
\hline\noalign{\smallskip}
			$\mu$  & $\mbox{Re}[\omega_{1}]$ & $\mbox{Im}[\omega_{1}]$ \\
\noalign{\smallskip}\hline\noalign{\smallskip}
			0 & 0.00000 & 15.29571 \\
			1 & 0.00000 & 16.04227 \\
			2 & 0.00000 & 17.91557 \\
			3 & 0.00000 & 20.37319 \\
			4 & 0.00000 & 23.11373 \\
			5 & 0.00000 & 26.00114 \\
			6 & 0.00000 & 28.97192 \\
			7 & 0.00000 & 31.99381 \\
\noalign{\smallskip}\hline
\end{tabular}
\end{table}

\newpage

\begin{table}[!ht]
\caption{The scalar resonant frequencies five-dimensional Schwarzschild black hole for $M=100$, $a=0.1$ and $\mu=5$.}
\label{tab:Tab3_Lovelock_D5}
\begin{tabular}{ccc}
\hline\noalign{\smallskip}
			$s$  & $\mbox{Re}[\omega_{7}]$ & $\mbox{Im}[\omega_{7}]$ \\
\noalign{\smallskip}\hline\noalign{\smallskip}
			0 & 0.00000 & 143.8524 \\
			1 & 0.00000 & 143.7013 \\
			2 & 0.00000 & 143.4441 \\
			3 & 0.00000 & 143.0720 \\
			4 & 0.00000 & 142.5702 \\
			5 & 0.00000 & 141.9147 \\
			6 & 0.00000 & 141.0635 \\
			7 & 0.00000 & 139.9352 \\
\noalign{\smallskip}\hline
\end{tabular}
\end{table}

\begin{table}[!ht]
\caption{The scalar resonant frequencies five-dimensional Schwarzschild black hole for $M=100$, $a=0.1$ and $\mu=5$.}
\label{tab:Tab4_Lovelock_D5}
\begin{tabular}{ccc}
\hline\noalign{\smallskip}
			$n$  & $\mbox{Re}[\omega_{n}]$ & $\mbox{Im}[\omega_{n}]$ \\
\noalign{\smallskip}\hline\noalign{\smallskip}
			0 & 0.00000 & 73.85206 \\
			1 & 0.00000 & 83.85211 \\
			2 & 0.00000 & 93.85217 \\
			3 & 0.00000 & 103.8522 \\
			4 & 0.00000 & 113.8523 \\
			5 & 0.00000 & 123.8523 \\
			6 & 0.00000 & 133.8524 \\
			7 & 0.00000 & 143.8524 \\
\noalign{\smallskip}\hline
\end{tabular}
\end{table}

From Table \ref{tab:Tab1_Lovelock_D5} we see that the (imaginary part of) resonant frequencies increase very slowly with $a$, for fixed values of the parameters $M$, $\mu$, $s$, and $n$. In Table \ref{tab:Tab2_Lovelock_D5} we can conclude that the (imaginary part of) resonant frequencies increase with $\mu$, for fixed values of the parameters $M$, $a$, $s$, and $n$.

In Table \ref{tab:Tab3_Lovelock_D5}, the magnitude of the resonant frequencies decreases with parameter $s$, for fixed values of the parameters $M$, $a$, $\mu$, and $n$. Finally, from Table \ref{tab:Tab4_Lovelock_D5}, we conclude that the magnitude of the resonant frequencies increase very quickly with parameter $n$, for fixed values of the parameters $M$, $a$, $\mu$, and $s$.
%
%%%%%%%%%%%%%%%%%%%%%%%%%%%%%%%%%%%%%%%%%%%%%%%%%%%%%%%%%%%%%%%%%%%%%%%%%%%%%%%%%%%%%%%%%%%%%% Conclusions
%
\section{Summary}\label{Conclusion}
In this work we have presented the analytical solutions of the both angular and radial parts of the Klein--Gordon equation in the five-dimensional Schwarzschild black hole spacetime in a string cloud model. In what concern the angular part, its general solution is given in terms of the three-dimensional normalized spherical harmonic function. On the other hand, for the radial part, its general solution is given in terms of the general Heun functions.

From the radial solution, we studied two very important physical phenomena. We get the Hawking radiation spectrum, which is similar to the black body radiation, for massive scalar fields nearby to the exterior apparent horizon. We obtained an expression for the massive scalar resonant frequencies, by imposing the appropriated boundary condition, and then examined their behavior as functions of the involved parameters.

These two physical processes, which are related to the interaction between quantum fields, in special the scalar one, and the gravitational field of black holes, in particular the five-dimension Lovelock spacetime, can give us, in principle, some relevant information about the physics of these objects. Therefore, they constitute a very important line of research and then need to be investigated in both theoretical and experimental point of view. Here we have presented some analytical and numerical results which can be compared with future detected data.

Finally, it is worth commenting that in addition to the calculation of the relative
scattering probability (or tunneling rate), which is given by
Eq.~\ref{eq:taxa_refl_Lovelock_D5}, the canonical invariance and a temporal
contribution to the tunneling rate can be also discussed, and therefore, it would be
interesting to extend our analysis to some other stationary and non-stationary black
hole spacetimes. In fact, some results which concern this extension were already
published~\cite{PhysLettB.642.124,IntJModPhysA.22.1705,Pramana.70.3,IntJModPhysD.17.2453}
and we expect to publish some others in the near future.
%
%%%%%%%%%%%%%%%%%%%%%%%%%%%%%%%%%%%%%%%%%%%%%%%%%%%%%%%%%%%%%%%%%%%%%%%%%%%%%%%%%%%%%%%%%%%%%% acknowledgments
%
\section*{Acknowledgement}
H.S.V. is funded by the Coordena\c c\~{a}o de A\-per\-fei\-\c co\-a\-men\-to de Pessoal de N\'{i}vel Superior - Brasil (CAPES) - Finance Code 001. The author also would like to thank Professor Daniel Vanzella for the very fruitful discussions.
%
%%%%%%%%%%%%%%%%%%%%%%%%%%%%%%%%%%%%%%%%%%%%%%%%%%%%%%%%%%%%%%%%%%%%%%%%%%%%%%%%%%%%%%%%%%%%%% thebibliography
%

%
%%%%%%%%%%%%%%%%%%%%%%%%%%%%%%%%%%%%%%%%%%%%%%%%%%%%%%%%%%%%%%%%%%%%%%%%%%%%%%%%%%%%%%%%%%%%%%
%
\end{document}